\documentstyle[aps,epsf,prl, multicol]{revtex}

\begin{document}
\title{ Nonlinear  Decoherence in Quantum State Preparation\\ of a Trapped Ion }
\author{Le-Man Kuang, Hao-Sheng Zeng, and Zhao-Yang Tong}
\address{ Department of Physics, Hunan Normal University, Changsha 410081,
             China}
\maketitle
\begin{abstract}
We present a nonlinear decoherence model which models decoherence
effect caused by various decohereing sources in a  quantum system
through  a nonlinear coupling between  the system and its
environment,  and apply it to  investigating decoherence in
nonclassical motional states of a single trapped ion.  We obtain
an exactly analytic solution of the model and  find very good
agreement with experimental results for the population decay rate
of a single trapped ion observed in the NIST experiments by
Meekhof and coworkers (D. M. Meekhof, {\it et al.}, Phys. Rev.
Lett. {\bf 76}, 1796 (1996)).

\end{abstract}


\begin{multicols}{2}
 In recent years, much progress has been made in preparation, manipulation, and
measurement of quantum states of the center-of-mass vibrational
motion of a single trapped ion experimentally [1-8] and
theoretically [9-16], which are not only of fundamental physical
interest but also of practical use for sensitive detection of weak
signals [17] and quantum computation in an ion trap [3,9]. In
particular, the NIST group [4]  has experimentally created and
observed nonclassical motional states of a single trapped ion. In
the NIST experiments [4], an anti Jaynes-Cummings model (JCM)
interaction  between the internal and motional states of a trapped
ion is realized through  stimulated Raman transitions, which
couple internal states of the trapped ion to its  motional states,
when the Lamb-Dicke limit is satisfied and the driving laser
fields are tuned to the first blue sideband. Detection of motional
states is carried  out by observing the evolution characteristics
of quantum dynamics  of internal  levels of the trapped ion under
influence of the anti JCM-typed interaction. The NIST experiments
revealed  the fact that the population of the low atomic state
($P_{\downarrow}$) evolves according to the following
phenomenological expression
\begin{equation}
P_{\downarrow}(t)=\frac{1}{2}\{1+ \sum_n p_n\cos(2gt\sqrt{n+1})e^{-\gamma_nt}\}
\end{equation}
where $p_n$ is the initial probability distribution of motional states of the
trapped ion in the Fock representation, $g$ is a coupling constant between the
atomic internal and motional states, $\gamma_n$ is a decay rate. The experimentally
observed decay rate is of the following form
\begin{equation}
\gamma_n=\gamma_0(n+1)^{\nu}
\end{equation}
where the observed value of $\nu$ is $\nu\doteq 0.7 $.

A question that naturally arises is: how to explain the above
experimentally observed decay rate? It is generally accepted that
the appearance of the decay factor $\gamma_n$ in the  evolution of
internal states is a consequence of decoherence. It is of
practical significance to well understand decoherence for
preparation of nonclassical states and quantum computation in ion
traps.  There are various  sources of decoherence [1], such as ion
vibrational decoherence, ion internal-state decoherence,
decoherence caused by non-ideal external fields, and so on.
Recently, Schneider and Milburn [18] have investigated decoherence
due to  laser intensity and phase fluctuations and obtained the
power $\nu$ in Eq.(2) being $\nu\doteq 0.5$ instead of the
experimentally observed value $ 0.7 $.  More recently, Murao and
Knight [19], using master equation method, have studied
decoherence due to the imperfect dipole transitions and
fluctuation of vibrational potential in the NIST experiments. In
spite of these  efforts, the problem of decoherence in quantum
state preparation of a trapped ion has been not satisfactorily
solved, and its character and microscopic origin still call for
further attention.  In particular, it should be pointed out that 
the experimentally observed decay rate indicated in (2)  is a collective
 effect caused by various decohering sources, not by a specific decohering source. 
 Nevertheless, authors in refs[18,19] investigated the decay rate caused 
 only by a specific source  of decoherence, not by various sources of decoherence. 
 So  how to model the experimentally observed decay rate caused by various decohering 
 sources is an interesting subject in quantum state preparation and manipulation  
 of a trapped ion. 
In this paper,  we present a nonlinear decoherence
model to model decoherence effects caused by  various decohering sources in a quantum system. 
 We shall show that our theoretical model can well describe the experimentally 
observed decay rate in the NIST experiments [4].

 We consider a single trapped ion with mass $m$ and laser cooled to the
Lamb-Dicke limit. Following symbols Ref.[19], we denote three
related internal states and motional states of the ion by $|i
\rangle$ ($i=\underline{0}, \downarrow, \uparrow$) and $|n\rangle$
($n=0, 1, 2, ...$), respectively. The free Hamiltonian of the
trapped ion is given by $\hat{H_0}=\hbar\omega_x\hat{a}^+\hat{a} -
\hbar\omega_{01}|\downarrow\rangle \langle \downarrow|  -
\hbar\omega_{02}|\uparrow\rangle \langle  \uparrow|$, where
$\omega_{01} (\omega_{02})$ is the transition frequency between
states $|\downarrow\rangle  (|\uparrow\rangle )$ and
$|\underline{0}\rangle $, $\hat{a}^+ (\hat{a})$ is the creation
(annihilation) operator of the motional states with the
corresponding frequency $\omega_x$. Two driving laser beams with
detuning $\Delta$, wave vector $\vec{k}_1 (\vec{k}_2)$ and
frequency $\omega_1(\omega_2)$ are used to  cause dipole
transitions between the level  $|\downarrow\rangle
(|\uparrow\rangle )$ and $|\underline{0}\rangle $.

With the dipole and rotating wave approximations, under large detuning condition
the intermediate level $|\underline{0}\rangle $ can be adiabatically eliminated when the
Lamb-Dicke limit is met and the driving laser beam is tuned to the first blue
sideband. Then, in the interaction picture of $\hat{H_0}$, the effective Hamiltonian
of the system has the anti JCM-typed  form
\begin{equation}
\hat{H}_S=\hbar g(\hat{a}^+\sigma_+ + \hat{a}\sigma_-)
\end{equation}
where $g$ is a coupling constant, which depends on  the  coupling
strength between internal and motional states of the trapped ion
and the Lamb-Dicke parameter defined by $\eta=\delta kx_0$, where
$\delta k$ is the wave-vector difference of the two Raman beams
along $x$, and $x_0=\sqrt{\hbar/2m\omega_x}$. For simplicity, we
set $\hbar=1$ throughout this paper.

The Hamiltonian (3) is diagonal in the dressed-state representation with the
following basis
\begin{equation}
|\varphi(n,i)\rangle =\frac{1}{\sqrt{2}}(|\downarrow, n\rangle  - (-1)^i|\uparrow, n+1\rangle ), i=1,2
\end{equation}

\begin{equation}
|\varphi(0,3)\rangle =|\uparrow, 0\rangle
\end{equation}
And we have $\hat{H}_S|\varphi(n,i)\rangle =E_{ni}|\varphi(n,i)\rangle $
 with eigenvalues $E_{ni}=(-1)^{i+1}g\sqrt{n+1}$ for $i=1,2$, and $E_{03}=0$.

Before going to our model, let us briefly recall a few basic facts
about the interaction between a quantum system and its
environment. The interaction  between the  system and its
environment  may create two types of effects [20-34]: decoherence
and  dissipation, which can be mathematically described by
decaying of the off-diagonal and diagonal elements of the reduced
density operator of the system, respectively. These  two effects
have been paid much attention in various areas, for instance,
quantum measurement [20,25-28], condensed matter physics [21-23],
quantum computation [29-31], and so on. The decoherence effect
causes the states of the system continuously decohere  to approach
classical states [20,27]. The dissipation effect   dissipates
energy of the system to environment [21-23]. The two effects can
be understood in terms of Hamiltonian formalism [32-34]. If we
assume that the total Hamiltonian of the system plus environment
to be $\hat{H}_T=\hat{H}_S+\hat{H}_R+\hat{H}_I$, where $\hat{H}_S$
and $\hat{H}_R$ are Hamiltonians of the system and environment,
respectively, and $\hat{H}_I$ is the interaction Hamiltonian
between them,   when the  Hamiltonian of the system commutes with
that of the interaction  between the system and environment, i.e.,
$[\hat{H}_S, \hat{H}_I]=0$, which means that there is no energy
transfer between the system and the environment, energy of the
system is conservative, so that what interaction between the
system and environment describes is decoherence effect. When
$[\hat{H}_S, \hat{H}_I]\neq 0$, there is energy transfer between
the system and environment, so that  what interaction between the
system  and environment describes is the dissipation effect. It
should be pointed out that the decoherence and dissipation happen
at different time scales [29,30]. The dissipation  effect occurs
at the relaxation time $\tau_{rel}$, while the decoherence time
scale $\tau_d$  is much shorter  than  $\tau_{rel}$ with the time
evolution of a quantum system. Hence, we here restrict  our
attention on decoherence effect.

We now present our model.  We use a reservoir consisting of an
infinite set of harmonic oscillators to model the environment of
the single trapped ion in the NIST experiments, and assume that in
the interaction picture of $\hat{H}_0$  the total Hamiltonian is
of the following phenomenological form
\begin{eqnarray}
\hat{H}_T&=&\hat{H}_S + \sum_k\omega_k\hat{b}^{\dagger}_k\hat{b}_k
+ F(\{\hat{O}_S\})\sum_kc_k(\hat{b}^{\dagger}_k+\hat{b}_k)
\nonumber \\
&&+F^2(\{\hat{O}_S\})\sum_k\frac{c_k^2}{\omega_k^2}.
\end{eqnarray}
Here the first term is the Hamiltonian of the system in the
interaction picture given by Eq.(3); the second term is the
Hamiltonian of the reservoir; the third one represents the
interaction between the system and the reservoir with a coupling
constant $c_k$, where $\{\hat{O}_S\}$ is a set of linear operators
of the system or their linear combinations  in the same picture as
that of $\hat{H}_S$,  $F(\{\hat{O}_S\})$ is an operator function of
$\{\hat{O}_S\}$. In order to enable  what the interaction between
the system and the reservoir describes in Eq.(6) is decoherence
not dissipation, we require that the linear operator $\hat{O}_S$
commutes with the Hamiltonian of the system, i.e., $[\hat{O}_S,
\hat{H}_S]=0$.  It is well known that the decohering  process can  indeed 
be considered as a quantum measurement process. The conventional  
definition of a quantum measurement involves any form of interaction 
between a quantum object and a classical system.  Therefore, the interaction function 
$F(\{\hat{O}_S\})$ in the model (6) can involve any form of interaction between 
the system and environment. This enables it to model the collective decohering
 behavior caused by various decohering sources.
The concrete form of the function
$F(\{\hat{O}_S\})$ may be regarded as an experimentally determined
quantity. The last term in Eq.(6) is a renormalization term, which
is discussed in Ref.[21]. When $F(\{\hat{O}_S\})$ is a  linear and
nonlinear function of the linear operator  $\hat{O}_S$,  We call
decoherence described by the interaction  between the system and
the reservoir linear and nonlinear decoherence, respectively, in
the similar sense of the linear and nonlinear dissipation implied
in Ref.[21]. In this sense, decoherence investigated in Ref.[19]
is a kind of linear decoherence. In what follow we  shall show
that nonlinear decoherence can better describe the decay rate in 
 the the NIST experiments.

The Hamiltonian (6) can be exactly solved by making use of the following
unitary transformation
\begin{equation}
\hat{U}=\exp[F(\{\hat{O}_S\})\sum_k\frac{c_k}{\omega_k}(\hat{b}^{\dagger}_k-\hat{b}_k)].
\end{equation}

After applying the unitary transformation (7) to the total Hamiltonian
(6), we get a decoupled Hamiltonian $\hat{H}'_T=\hat{H}_S +
\sum_k\omega_k\hat{n}_k$, where $\hat{n}_k=\hat{b}^{\dagger}_k\hat{b}_k$.
 The density operator associated with the decoplued Hamiltonian is given by
\begin{equation}
\hat{\rho}'_T(t)=e^{-i\hat{H}'_Tt}\hat{\rho}'_T(0)e^{i\hat{H}'_Tt}
\end{equation}
where $\hat{\rho}'_T(0)=\hat{U}\hat{\rho}_T(0)\hat{U}^{-1}$, with
$\hat{\rho}_T(0)$  being  the initial total density operator. Through a
converse transformation of (7), it is straightforward to obtain the total
density operator associated with the original  Hamiltonian (6) with the
following expression
\begin{eqnarray}
\hat{\rho}_T(t)&=&e^{-i\hat{H}_St}\hat{U}^{-1}e^{-it\sum_k\omega_k\hat{n}_k}
\hat{U}\hat{\rho}_T(0) \nonumber \\
&&\times \hat{U}^{-1} e^{it\sum_k\omega_k\hat{n}_k}\hat{U}e^{i\hat{H}_St}.
\end{eqnarray}

We assume that the system and  reservoir are initially in thermal
equilibrium and  uncorrelated, so that
$\hat{\rho}_T(0)=\hat{\rho}_S(0)\otimes\hat{\rho}_R(0)$, where
$\hat{\rho}_S(0)$ and $\hat{\rho}_R(0)$  are the initial  density
operator of the system  and the reservoir, respectively.
$\hat{\rho}_R(0)$ can be expressed  as
$\hat{\rho}_R=\prod_k\hat{\rho}_k(0)$ where
$\hat{\rho}_k(0)=(1-e^{-\beta\omega_k})e^{-\beta\omega_k\hat{n}_k}$
is the density  operator of the $k$-th harmonic oscillator in
thermal equilibrium, where  $\beta=1/k_BT$,  $k_B$ and $T$ being
the Boltzmann constant and temperature, respectively.  After
taking the  trace over the reservoir, from  Eq.(9) we can get the
reduced density operator of the system, denoted by
$\hat{\rho}(t)=tr_R\hat{\rho}_T(t)$, its matrix elements in the
dressed state  representation
 are explicitly written as
\begin{eqnarray}
\rho_{(m',i')(m,i)}(t)&=&\rho_{(m',i')(m,i)}(0)R_{m'i'mi}(t)
\nonumber \\
&&\times e^{-i\phi_{m'i'mi}(t)},
\end{eqnarray}
Here the phase is defined by 
\begin{equation}
\phi_{m'i'mi}(t)=[E_{m'i'}-E_{mi}], 
\end{equation} 
and  $R_{m'i'mi}(t)$ is a reservoir-dependent quantity given by
\begin{eqnarray}
R_{m'i'mi}(t)&=&\prod_k
Tr_R\{D(-\alpha_{mik})e^{-it\omega_k\hat{n}_k}D(-\alpha_{mik})\nonumber \\
&&\times D(-\alpha_{m'i'k})e^{-it\omega_k\hat{n}_k}D(-\alpha_{m'i'k})
\hat{\rho}_k(0)\},
\end{eqnarray}
where $\alpha_{mik}=f(\{O_{mi}\})c_k/\omega_k$ with $O_{mi}$ being
an eigenvalue of the linear operator $\hat{O}_S$ in a dressed
state, i.e.,
$\hat{O}_S|\varphi(m,i)\rangle=O_{mi}|\varphi(m,i)\rangle$, and
$D(\alpha)=\exp(\alpha\hat{b}_k^+-\alpha^*\hat{b}_k)$ is a
displacement operator.

Making use of properties of the displacement operator:
\begin{eqnarray}
D(\alpha)D(\beta)=D(\alpha+\beta)\exp[iIm(\alpha\beta^*)], \\
 \exp(x\hat{n}_k)D(\alpha)\exp(-x\hat{n}_k)=
\exp(\alpha e^x\hat{b}_k^+-\alpha^* e^{-x}\hat{b}_k), 
\end{eqnarray}
and the following formula [35]
\begin{equation}
Tr_R[D(\alpha)\hat{\rho}_k(0)]=\exp[-\frac{1}{2}|\alpha|^2
\coth(\frac{\beta\omega_k}{2})],
\end{equation}
we find that the reservoir-dependent quantity $R_{m'i'mi}(t)$ can be
written as the following factorized form
\begin{equation}
R_{m'i'mi}(t)=e^{-i\delta\phi_{m'i'mi}(t)}e^{-\Gamma_{m'i'mi}(t)},
\end{equation}
with the following phase shift and damping factor 
\begin{equation}
\delta\phi_{m'i'mi}(t)=[F^2(\{O_{m'i'}\})-F^2(\{O_{mi}\})]Q_1(t),
\end{equation}

\begin{equation}
\Gamma_{m'i'mi}(t)=[F(\{O_{m'i'}\})-F(\{O_{mi}\})]^2Q_2(t).
\end{equation}
Here the two reservoir-dependent functions are given by
\begin{equation}
Q_1(t)=\int^{\infty}_{0} d\omega J(\omega)\frac{c^2(\omega)}{\omega^2}\sin(\omega t),
\end{equation}

\begin{equation}
Q_2(t)=2\int^{\infty}_{0} d\omega J(\omega)
\frac{c^2(\omega)}{\omega^2}\sin^2(\frac{\omega t}{2})\coth(\frac{\beta\omega}{2}),
\end{equation}
where we have taken the continuum limit of the reservoir modes:
$\sum_k \rightarrow \int^{\infty}_{0} d\omega J(\omega)$,  where
$J(\omega)$ is the spectral density of the reservoir, $c(\omega)$
is the corresponding continuum expression for $c_k$.

We assume that the system is initially in a state
$\hat{\rho}(0)=|\downarrow\rangle \langle  \downarrow|\otimes\sum_np_n|n
\rangle\langle  n|$. Then, from
Eqs.(10)-(16) we find that at time $t$ the population of the lower atomic state is
given by
\begin{equation}
P_{\downarrow}(t)=\frac{1}{2}\{1+ \sum_n
p_n\cos[\phi_{n1n2}(t)+\delta\phi_{n1n2}(t)]e^{-\Gamma_{n1n2}(t)},
\end{equation}
which indicates that the interaction between the system and reservoir induces a
phase shift $\delta\phi_{n1n2}(t)$ and a damping factor $\Gamma_{n1n2}(t)$ in the
time evolution of the atomic population.

Taking into account the experimental expression (1), we  choose
the following  linear operator  and interaction function:
\begin{eqnarray}
\hat{O}_S=\hat{a}^+\sigma_+ + \hat{a}\sigma_-, \\
 F(\{\hat{O}_S\})=\hat{O}_S^{2d+1}, 
\end{eqnarray}
where $d$ is an adjustable
parameter to describe the nonlinearity in the interaction, which
reflects the deviation degree of the nonlinearity of
$F(\{\hat{O}_S\})$ with respect to the linear operator
$\hat{O}_S$. The value of the parameter $d$ is determined by the
experimental results. With these choices, it is easy to find that 
\begin{eqnarray}
F^2(O_{n1})-F^2(O_{n2})=0, \\
 F(O_{n1})-F(O_{n2})=2(\sqrt{n+1})^{2d+1}.
\end{eqnarray}
 Then the phase shift in Eq.(21) naturally  vanishes,  and the damping factor becomes
\begin{equation}
\Gamma_{n1n2}=4(n+1)^{2d+1}Q_2(t).
\end{equation}
  So that we can find from Eq.(21) that
\begin{equation}
P_{\downarrow}(t)=\frac{1}{2}\{1+ \sum_n
p_n\cos(2gt\sqrt{n+1})e^{-4(n+1)^{\nu}Q_2(t)}\}
\end{equation}
where  $\nu =2d+1$ and $Q_2(t)$ is given by Eq.(20). From Eq.(27)
we see  that the argument of the cosine function on the RHS of
Eq.(27) does have  the   same form as that  in the experimental
expression (1).
 Comparing the theoretical expression (27) with the experimental result (1), we find
that when the nonlinear deviation $d\doteq -0.15$, the $n$-dependence of the damping
factor in Eq.(27) is completely in agreement  with that  seen in the experimental
expression (1). The final step is to determine the time dependence of the damping
factor in Eq.(27). From Eqs.(19), (20), and (27), we see that   all necessary
information about the effects of the environment  is contained in the spectral
density of the reservoir.  Eq.(27) indicates that the time dependence of the damping
factor is completely determined by the  spectral density of the reservoir.
The experimental expression (1) requires that the time dependence of the damping
factor must be linear, so that if we choose the spectral density such that
\begin{equation}
Q_2(t)=\frac{1}{4}\gamma_0t
\end{equation}
where $\gamma_0$ is a characteristic parameter, then we can get an
expression of $P_{\downarrow}(t)$, which  has exactly the same
form as the experimental result (1).   It is possible to find a
spectral density of the reservoir to satisfy the condition (28).
For instance, for the case of zero temperature, if we   take the
spectral density  $J(\omega)=\gamma_0/(2\pi c^2(\omega))$,
substituting it to Eq.(20) we can realize  Eq.(28).

In conclusion, we have present a nonlinear decoherence model, and
obtained its exactly analytic solution. It has been shown that our
model can give precisely the same expression of the population
decay rate of the single trapped ion  as  that observed in the
NIST experiments [4].  The nonlinear decoherence model can
describe the NIST experiments so well. This indicates that the
reservior and the nonlinear coupling  between the system and the
reservoir, which we design, properly model the real environment of
the single trapped ion in the experiments. 
It is worthwhile to emphasize that the nonlinearity in the
coupling describes a collective contribution of various decohering
sources to the decay rate. Hence, what the
nonlinear decoherence  describes is a collective decoherence
effect caused by various decohering sources not a specific
decoherence source.  We have noted that authors in ref.[19] obtained 
the decay rate in Eq.(1), but decoherence which they considered is a 
specific decoherence  caused by the imperfect dipole transitions and
fluctuation of vibrational potential, so their results can not cover 
the contribution of other decohering sources to the decay rate in the
 NIST experiment. It can  be expected that the nonlinear
decoherence model proposed in the present paper  can describe
decoherence behaviors of a wide variety of quantum systems.

L. M. Kuang thanks Profs.  Changpu Sun , Hong Chen , and  Dr. Shao-Ming Fei a for 
enlightening discussions. This work was supported in part
by 95-climbing project of China,  NSF of China,      Educational Committee Foundation and NSF of Hunan
Province, and  special project of NSF of China via Institute of
Theoreical Physics, Academia Sinica.

\end{multicols}
\end{document}